\documentclass[aip,graphicx]{revtex4-1}

\usepackage{graphicx}
\usepackage{dcolumn}
\usepackage{bm}

\usepackage[utf8]{inputenc}
\usepackage[T1]{fontenc}
\usepackage{mathptmx}

\begin{document}

\title[Properties of Grooved Dayem Bridge based YBa$_2$Cu$_3$O$_{7-\delta}$ SQUIDs and Magnetometers]{Properties of Grooved Dayem Bridge based YBa$_2$Cu$_3$O$_{7-\delta}$ Superconducting Quantum Interference Devices and Magnetometers}

\author{E. Trabaldo}
\affiliation{Quantum Device Physics Laboratory, Department of Microtechnology and Nanoscience, Chalmers University of Technology, SE-41296 G\"{o}teborg, Sweden}
\author{S. Ruffieux}%
\affiliation{Quantum Device Physics Laboratory, Department of Microtechnology and Nanoscience, Chalmers University of Technology, SE-41296 G\"{o}teborg, Sweden}
\author{E. Andersson}
\affiliation{Quantum Device Physics Laboratory, Department of Microtechnology and Nanoscience, Chalmers University of Technology, SE-41296 G\"{o}teborg, Sweden}
\author{R. Arpaia}
\affiliation{Quantum Device Physics Laboratory, Department of Microtechnology and Nanoscience, Chalmers University of Technology, SE-41296 G\"{o}teborg, Sweden}
\affiliation{Dipartimento di Fisica, Politecnico di Milano, Piazza Leonardo da Vinci 32, I-20133 Milano, Italy}
\author{D. Montemurro}
\altaffiliation[Present address: ]
{Dipartimento di Fisica, Università di Napoli Federico II, Via Cinthia, I-80126 Napoli, Italy}
\affiliation{Quantum Device Physics Laboratory, Department of Microtechnology and Nanoscience, Chalmers University of Technology, SE-41296 G\"{o}teborg, Sweden}
\author{J.F. Schneiderman}
\affiliation{MedTech West and the Institute for Neuroscience and Physiology, Sahlgrenska Academy, University of Gothenburg, SE-40530 G\"{o}teborg, Sweden}
\author{A. Kalaboukhov}
\affiliation{Quantum Device Physics Laboratory, Department of Microtechnology and Nanoscience, Chalmers University of Technology, SE-41296 G\"{o}teborg, Sweden}
\author{D. Winkler}
\affiliation{Quantum Device Physics Laboratory, Department of Microtechnology and Nanoscience, Chalmers University of Technology, SE-41296 G\"{o}teborg, Sweden}
\author{F. Lombardi}
\affiliation{Quantum Device Physics Laboratory, Department of Microtechnology and Nanoscience, Chalmers University of Technology, SE-41296 G\"{o}teborg, Sweden}
\author{T. Bauch}
\email{thilo.bauch@chalmers.se}
\affiliation{Quantum Device Physics Laboratory, Department of Microtechnology and Nanoscience, Chalmers University of Technology, SE-41296 G\"{o}teborg, Sweden}

\date{\today}

\begin{abstract}
The transport properties of a YBa$_2$Cu$_3$O$_{7-\delta}$ superconducting quantum interference device (SQUID) based on grooved Dayem bridge weak links are studied as a function of temperature: at high temperatures ($60~$K$<T<T_\mathrm{c}=89$~K) the weak links show properties similar to SNS junctions, while at temperatures below 60~K the weak links behave like short Dayem bridges. Using these devices, we have fabricated SQUID  magnetometers with galvanically coupled in-plane pick-up loops: at $T=77$~K, magnetic field white noise levels as low as $63$~fT/$\sqrt{\mathrm{Hz}}$ have been achieved.
\end{abstract}

\maketitle

One of the most prominent applications of superconducting materials is the superconducting quantum interference device (SQUID)\cite{weinstock2012squid}. SQUIDs are used in geophysical surveys and mining, non-destructive structure evaluation, scanning SQUID microscopy, and biomagnetic  diagnostics (magnetoencephalography, magnetocardiography, etc) \cite{fagaly2006superconducting,clarke2006squid,seidel2015applied,clarke2018focus}. State of the art SQUIDs are based on low critical temperature superconductors (LTS)\cite{granata2016nano} typically requiring expensive and scarce liquid helium for their operation. High critical temperature superconductor (HTS) based SQUIDs, instead, can be  operated at liquid nitrogen temperatures. The use of cheap and abundant liquid nitrogen furthermore simplifies the cryogenic requirements in terms of cooling and thermal insulation as compared to LTS-based systems. The realization of HTS Josephson junctions, the key ingredient of a SQUID, has been a topic of intense research during the last three decades \cite{hilgenkamp2002grain,koelle1999high,martinez2016nanosquids,tafuri2005weak,cho2018superconducting}. The state-of-the-art HTS SQUIDs operating near 77~K typically use either bicrystal or step edge grain boundary Josephson junctions.\cite{faley2014graphoepitaxial,mitchell2010ybco,oisjoen2012high,nagel2010resistively,beyer1998low} However, bicrystal junctions need to be placed at the grain boundary line, while step edge junctions require more than one lithography step\cite{foley1999fabrication} and several epitaxial thin film depositions.\cite{faley2014graphoepitaxial} The recent development of the YBCO grooved Dayem bridge (GDB) made it possible to realize SQUID magnetometers with magnetic field noise values at 77~K comparable to the best single layer devices.\cite{trabaldo2019grooved} In contrast to grain boundary JJs, GDBs can be defined anywhere on the chip and oriented at will within the film plane. Moreover, the bridge and weak link inside it are realized during a single lithography process. However, the nature of the GDB weak link determining its transport properties has not been explored yet. 

In this letter we present a study of GDB-based SQUID properties as a function of temperature. From the temperature dependence of these properties, we conclude that the GDB is governed by superconductor-normal conductor-superconductor (SNS) like behaviour at $T>60~$K, whereas at lower temperatures the weak link is better described by an SS$^\prime$S type of weak link, where S$^\prime$ describes a superconducting constriction. By coupling a GDB-based SQUID galvanically to a 9~mm$~\times$~8.7~mm pick-up loop on a $10\mathrm{~mm}\times 10$~mm substrate we achieve a magnetic field sensitivity as low as 63~fT/$\sqrt{\mbox{Hz}}$ at 77~K.

\begin{figure}
\includegraphics[width=0.45\textwidth]{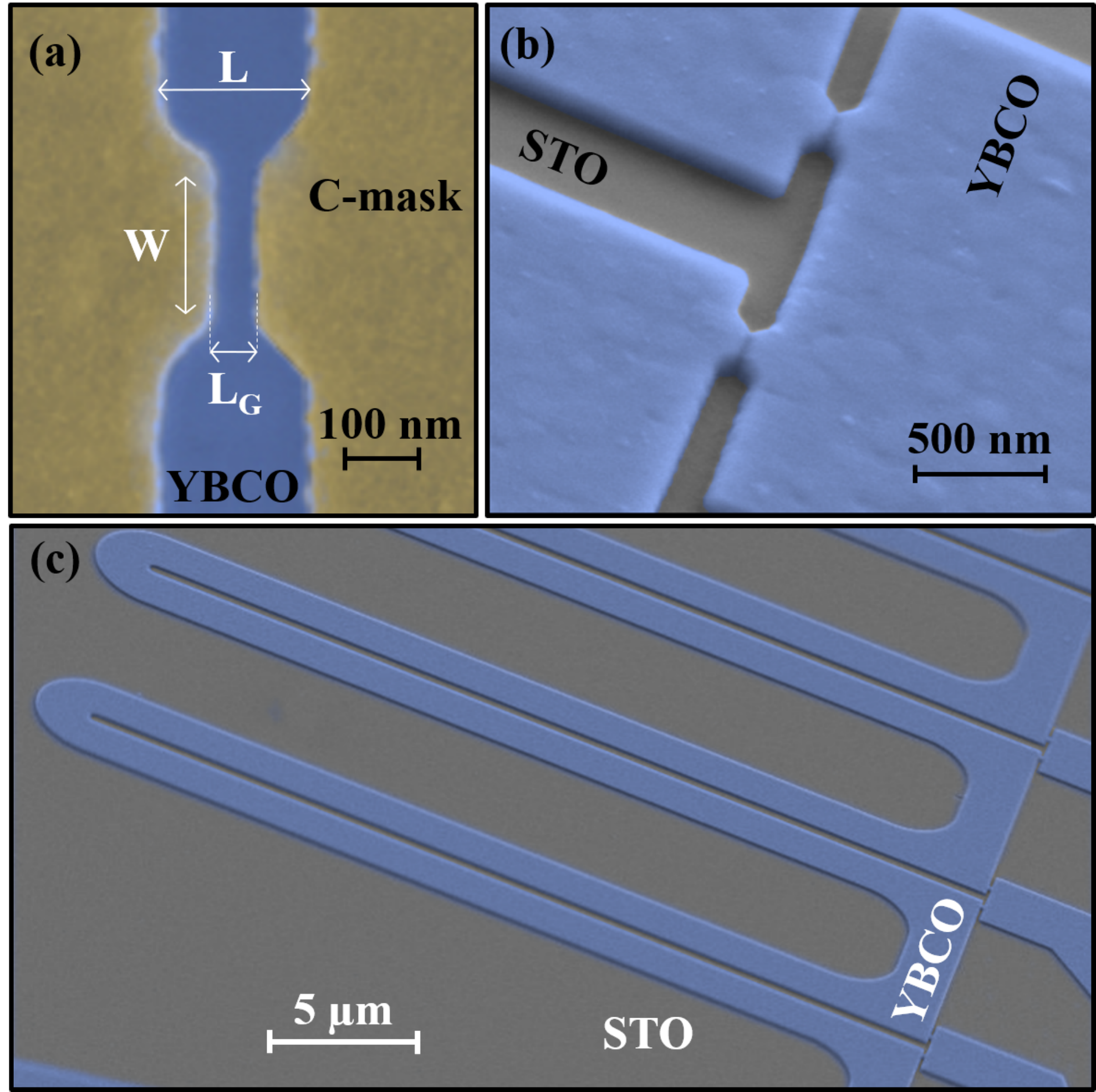}
\caption{\label{fig1}(a) Example of a carbon mask for the realization of a GDB prior to Ar$^+$ ion milling. The mask dimensions are $W=200$~nm, $L=150$~nm and $L_\mathrm{G}=50$~nm. (b) SEM image of two GDBs patterned in a SQUID loop. (c) SEM image of a series of SQUID loops.}
\end{figure}

The SQUIDs are fabricated by first depositing a $50$~nm thick YBa$_2$Cu$_3$O$_{7-\delta}$ (YBCO) film on a (001) oriented SrTiO$_3$ substrate by pulsed laser deposition (PLD). The lateral size of the substrate is $10\mathrm{~mm}\times 10$~mm which allows accommodation of a pick-up loop, and thus an effective area, that is significantly larger as compared to our previous work. \cite{trabaldo2019grooved} The deposition conditions have been tuned so that the YBCO film is close to optimally doped (critical temperature $T_\mathrm{c}\simeq89.0$~K) and $c$-axis oriented. \cite{arpaia2018probing, baghdadi2014toward} An amorphous carbon film, with thickness $t_\mathrm{C}=100$~nm, is deposited by PLD on top of the YBCO film, which is subsequently patterned by electron beam lithography (EBL) and oxygen plasma reactive ion etching. The pattern of the carbon mask is transferred to the YBCO film using low voltage (300~eV) Ar$^+$ ion milling. The Ar$^+$ ion milling parameters have been tuned to minimize damage to the superconducting nanostructures. \cite{nawaz2013approaching,arpaia2016improved,trabaldo2017noise} 

A Grooved Dayem Bridge (GDB) is obtained by designing a gap in the carbon mask, along the whole width of a short nanowire (Dayem bridge),  with a typical gap length $L_\mathrm{G}=50$~nm, see Fig.\ref{fig1}(a).
The etching rate during Ar$^+$ ion milling of the material inside the nano-gap is decreased compared to the rest of the sample, when $t_\mathrm{C}/L_\mathrm{G}\gtrsim2$. \cite{trabaldo2019grooved,manos1989plasma} In this regime, the increased re-deposition rate inside the nano-gap leads to a reduced effective etching rate. The final result of the Ar$^+$ ion milling is a nano-bridge with width $W=200$~nm and length $L=150$~nm and with a groove etched in the center, which acts as a weak link (see Fig.\ref{fig1}(b)). The whole mask design (nanostructures and pick-up loop) is defined in a single EBL step, avoiding alignment errors, which are common in nano-fabrication processes involving several lithography steps. \cite{baghdadi2015fabricating,granata2013three} The single step of Ar$^+$ ion milling required to define the device simplifies the fabrication process and minimizes the detrimental effect of Ar$^+$ ion milling to the YBCO nanostructures. The SQUID loop has a hairpin design \cite{xie2017improved} (see Fig.~\ref{fig1}(c)) with hair pin slit length, $l_\mathrm{slit}$, ranging from $8$ to $36$~$\mu$m, slit width of $500$~nm, and line width of $2$~$\mu$m. The GDBs act as the SQUID weak links (Fig.~\ref{fig1}(b)) where the groove locally reduces the $I_\mathrm{c}$ of the nano-bridge. The electric and magnetic parameters of the 4 SQUIDs studied in this work are summarized in Table~\ref{tab1}.

The sensitivity of a SQUID magnetometer is limited by its magnetic flux noise $S^{1/2}_\Phi$ (1/f and white), and its effective area, $A_\mathrm{eff}$. The minimal intrinsic flux noise of a SQUID can be achieved for a screening parameter \cite{tesche1977dc} $\beta_\mathrm{L}=I_\mathrm{c}L_\mathrm{SQ}/\Phi_0$ close to 1. Here, $I_\mathrm{c}$ is the SQUID critical current, $L_\mathrm{SQ}$ is the total inductance of the SQUID (including hair pin loop inductance $L_\mathrm{c}$ and the parasitic kinetic  inductance of the GDB $L_k$), and $\Phi_0$ the superconducting flux quantum. The large critical current density, $j_\mathrm{c}\simeq 3\cdot 10^6~$A/cm$^2$, at 77~K of bare YBCO Dayem bridges \cite{nawaz2013approaching,arpaia2014ultra}, together with the parasitic kinetic inductance of a Dayem bridge of thickness $t$ and width $w$, $L_k = L\mu_0\lambda^2/wt$, sets a lower bound for the screening factor, $2 I_\mathrm{c} L_\mathrm{k}/\Phi_0 =  4j_\mathrm{c} L\mu_0 \lambda^2 /\Phi_0$ on the order of 1-3 at 77~K for a typical bridge length $L=  150~$nm.\cite{arpaia2014ultra} Here $\lambda$, and $\mu_0$ are the London penetration depth, and the vacuum permeability, respectively. Instead, for SQUIDs implementing GDBs, the contribution of the parasitic kinetic inductance to $\beta_\mathrm{L}$ is well below 0.3 at $T=77~$K, because the critical current density, $I_c^{G}/wt$, with $I_{c}^{G}$ the critical current of the GDB, is at least a factor of 10 lower than that of bare Dayem bridges.\cite{trabaldo2019grooved} This furthermore allows for sizable SQUID hair pin loops, which increases the effective area of the magnetometer and minimizes the resulting magnetic field noise as will be discussed below. 

\begin{figure}
\includegraphics[width=0.42\textwidth]{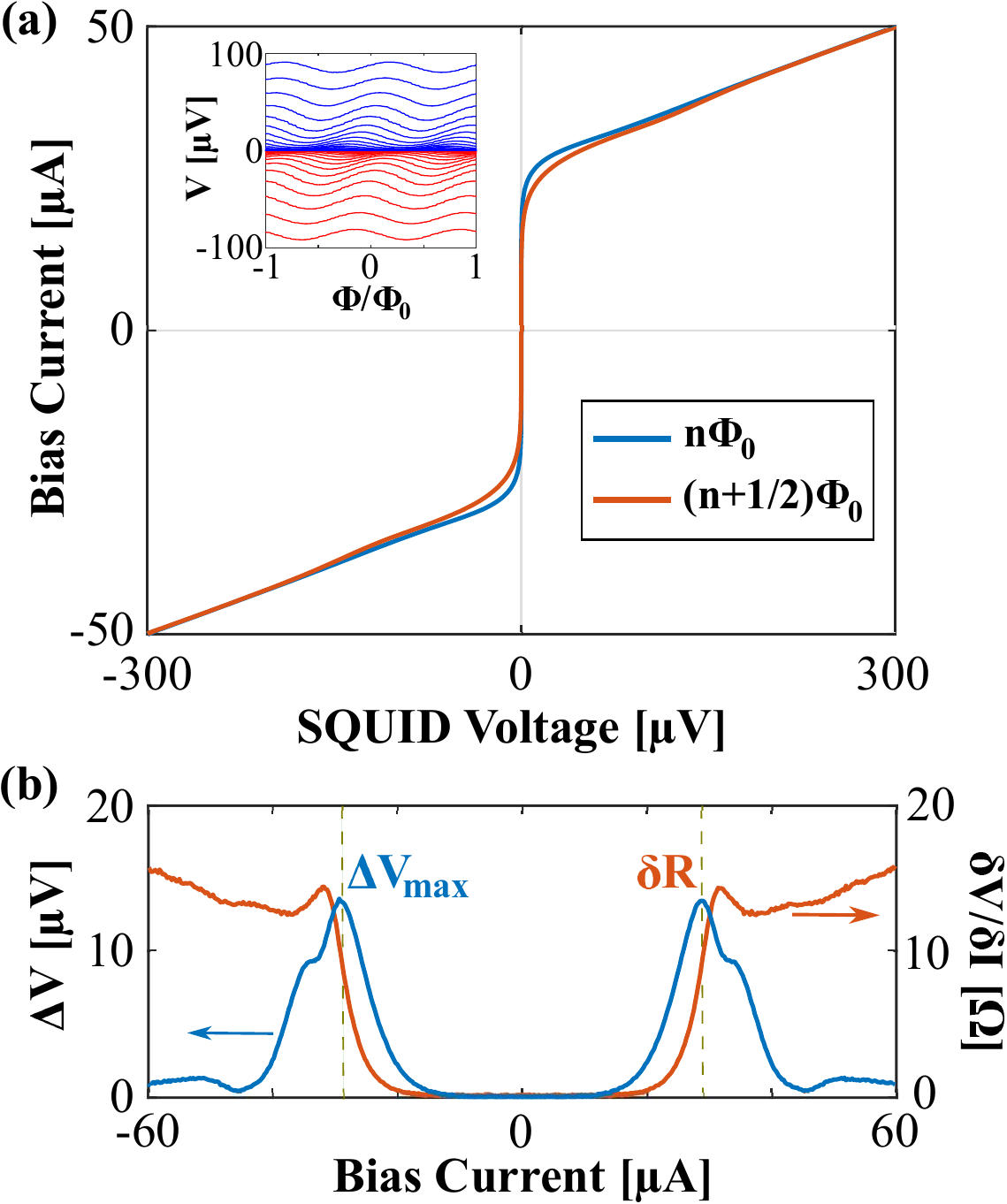}
\caption{\label{fig2} (a) Current voltage characteristic of SQUID SQ2 at $T=77$~K for two different values of applied magnetic flux. The inset shows the SQUID voltage modulations as a function of applied magnetic flux for various fixed bias currents with current step size $2~\mu$A (b) Differential resistance $\delta V/\delta I$ for $n\Phi_0$ and voltage modulation depth $\Delta V$ as a function of applied bias current $I_\mathrm{b}$. The value $\delta R$ is indicated as the differential resistance at the optimal working point where  $V_\Phi$ is maximized ($\Delta V = \Delta V_{\mathrm{max}}$).}
\end{figure}

The amplifier input voltage noise $S_\mathrm{V,a}^\mathrm{1/2}$ contribution to the total white flux noise should moreover be minimized. This contribution is given by $S_{\Phi \mathrm{,a}}^\mathrm{1/2}=S_\mathrm{V,a}^\mathrm{1/2}/V_\mathrm{\Phi}$, where 
$V_\Phi=\mathrm{max}(\delta V/\delta \Phi)\simeq \pi\Delta V_{\mathrm{max}}/\Phi_\mathrm{0}$ is the SQUID voltage-to-flux transfer function. $\Delta V_{\mathrm{max}}$ is the maximum voltage modulation depth in response to an externally applied magnetic flux (see Fig. \ref{fig2}).
$\Delta V_{\mathrm{max}}$ can be approximated by $\Delta I_{\mathrm{c}}\delta R$, where $\Delta I_{\mathrm{c}}$ and $\delta R$ are the critical current modulation depth and the differential resistance at the optimal bias current working point (i.e., where $\Delta V = \Delta V_{\mathrm{max}}$), respectively. Therefore, a large value of $\delta R$ is desirable in order to minimize the contribution of the amplifier voltage input noise to the total flux noise.

In Fig.~\ref{fig2}(a) we show the current voltage characteristics of SQUID SQ2 (see Table~\ref{tab1} for details) measured at $T=77$~K for two different applied magnetic flux values. The shape of the current voltage characteristics resembles that of a resistively shunted junction (RSJ). The voltage modulations of SQUID SQ2 as a function of applied magnetic flux are shown in the inset of Fig.~\ref{fig2}(a). Here each curve corresponds to an increment of bias current, $I_\mathrm{b}$, by $2$~$\mu$A. From this measurement one can extract the voltage modulation depth as a function of the bias current, which is shown in Fig.~\ref{fig2}(b). From the maximum voltage modulation depth $\Delta V_{\mathrm{max}}=16.5~\mu$V we obtain for the transfer function $V_{\Phi}=52$~$\mu\mathrm{V}/\Phi_0$ at $T=77$~K. This is a net improvement compared to nanowire-based SQUIDs with similar SQUID loop size $l_\mathrm{slit}$\cite{xie2017improved}, which is attributed to the increased differential resistance and reduced parasitic inductance of GDBs as compared to bare nanowires. As will be shown below, this results in lower magnetic flux and field noise as well. The values of $\Delta V_{\mathrm{max}}$ measured at $T=77~$K for the other SQUIDs are summarized in Table~\ref{tab1}.

\begin{figure*}
\includegraphics[width=\textwidth]{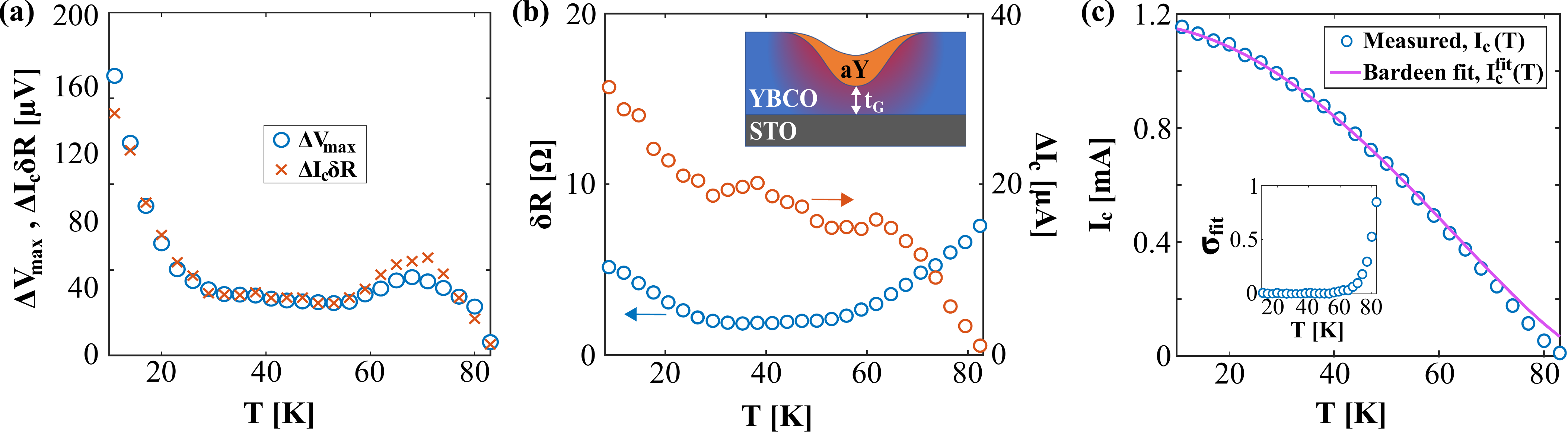}
\caption{\label{fig3} (a) The voltage modulation at working point $\Delta V_\mathrm{max}$ (cirles), obtained as the maximum of the measured $\Delta V$,  is compared as a function of the temperature to the product $\Delta I_\mathrm{c}\delta R$ (crosses). (b) Differential resistance at the working point $\delta R$ and critical current modulation depth $\Delta I_\mathrm{c}$ vs. $T$. The inset shows a schematic cross section of a GDB. $t_\mathrm{G}$ is the thickness of the superconducting YBCO in the groove, while aY represents the amorphous YBCO layer re-deposited during Ar$^+$ ion milling. (c) Measured critical current versus temperature, $I_\mathrm{c}$ (open symbols), compared to the Bardeen fit, $I^\mathrm{fit}_\mathrm{c}$ (solid line). Inset show the deviation, $\sigma _\mathrm{fit}$, of the measured critical current from the fit as $\sigma_\mathrm{fit}=(I^\mathrm{fit}_\mathrm{c}-I_\mathrm{c})/I^\mathrm{fit}_\mathrm{c}$}
\end{figure*}

In Fig.~\ref{fig3} (a), we show the maximum modulation depth $\Delta V_{\mathrm{max}}$ of SQUID SQ0 as a function of temperature $T$. In the temperature range between 10~K and 20~K,  $\Delta V_{\mathrm{max}}(T)$ decreases rapidly with temperature. Increasing the temperature above 20~K up to $T \sim 55$~K causes $\Delta V_{\mathrm{max}}$ to further decrease slightly. For temperatures above 55~K we observe an increase of $\Delta V_{\mathrm{max}}$ up to $T=65$~K above which the maximum voltage modulation depth goes to zero when reaching the critical temperature of the GDBs, $T^\mathrm{GDB}_\mathrm{c}\simeq 84$~K.  

The non-monotonic behavior of $\Delta V_{\mathrm{max}}(T)$ can be understood from the temperature dependence of the critical current modulation depth $\Delta I_{c}$ and the differential resistance $\delta R$ at the optimal working point, which are shown in Fig.~\ref{fig3}(b). The critical current modulation depth decreases with increasing temperature whereas the differential resistance first decreases with temperature until it increases again for temperatures above $T = 55~$K. The increase of the differential resistance with temperature is a typical feature of an SNS-like junction. The product $\Delta I_{\mathrm{C}} \delta R$, shown in comparison with $\Delta V_{\mathrm{max}}(T)$ in Fig.~\ref{fig3}(a) (crosses), reproduces nicely the maximum value of measured voltage modulation depth.

In Fig.~\ref{fig3} (c) we show the critical current $I_{\mathrm{c}}$ of device SQ0 as a function of temperature (blue circles). The solid line is the Bardeen expression for the depairing critical current of superconducting nanowires \cite{nawaz2013approaching,bardeen1962critical}: $I^\mathrm{fit}_\mathrm{c}(T)\propto (1-(T/T_\mathrm{c})^2)^{3/2}$. Here we used $T_{\mathrm{c}}=89~$K, the critical temperature of the YBCO film. In our previous work\cite{nawaz2013approaching} we have shown that the Bardeen expression properly reproduces the critical current of YBCO nanowires in the full temperature regime. The fact that the critical current of our GDBs can be well fitted by the Bardeen expression for temperatures below $T\simeq 60~$K suggests that the GDBs behave like short Dayem bridges at low temperatures. Here the length of the bridge is approximately given by the GDB gap length ($L_\mathrm{G}=50$~nm) and the thickness is given by the thickness of the remaining superconducting film, $t_\mathrm{G}$, inside the gap buried under the  redeposited amorphous YBCO \cite{trabaldo2019grooved} (see inset Fig.~\ref{fig3}(b)). From the typical critical current density of thin YBCO Dayem bridges\cite{arpaia2017transport} $j_{\mathrm{c}}\simeq 2\times 10^{7}~$A/cm$^2$ at 4.2~K, we therefore obtain for the thickness of the superconducting constriction in the GDB $t_\mathrm{G}\simeq 15$~nm.

For $T>60~$K the measured critical current, $I_\mathrm{c}(T)$, clearly departs from the temperature dependence of a bare Dayem bridge predicted by the Bardeen expression, $I^\mathrm{fit}_\mathrm{c}(T)$. A possible explanation is the weakening of superconductivity in the constriction of the GDB for $T$ approaching $T^\mathrm{GDB}_\mathrm{c}\simeq 84$~K, which could also explain the increase of the differential resistance above $T=60~$K. Indeed, in thin ($t\leq 30$~nm) YBCO films, a clear broadening of the superconducting transition in resistance vs temperature measurements has been observed \cite{arpaia2017transport}. This broadening can be attributed to a Kosterlitz-Thouless vortex-antivortex pair dissociation transition close to the $T_{\mathrm{c}}$ of the film\cite{beasley1979possibility,bartolf2010current}. Alternative explanations include thermal activation of vortex-antivortex pairs or vortices overcoming the Bean-Livingston edge barrier.\cite{arpaia2014resistive}

\begin{figure}
\includegraphics[width=0.48\textwidth]{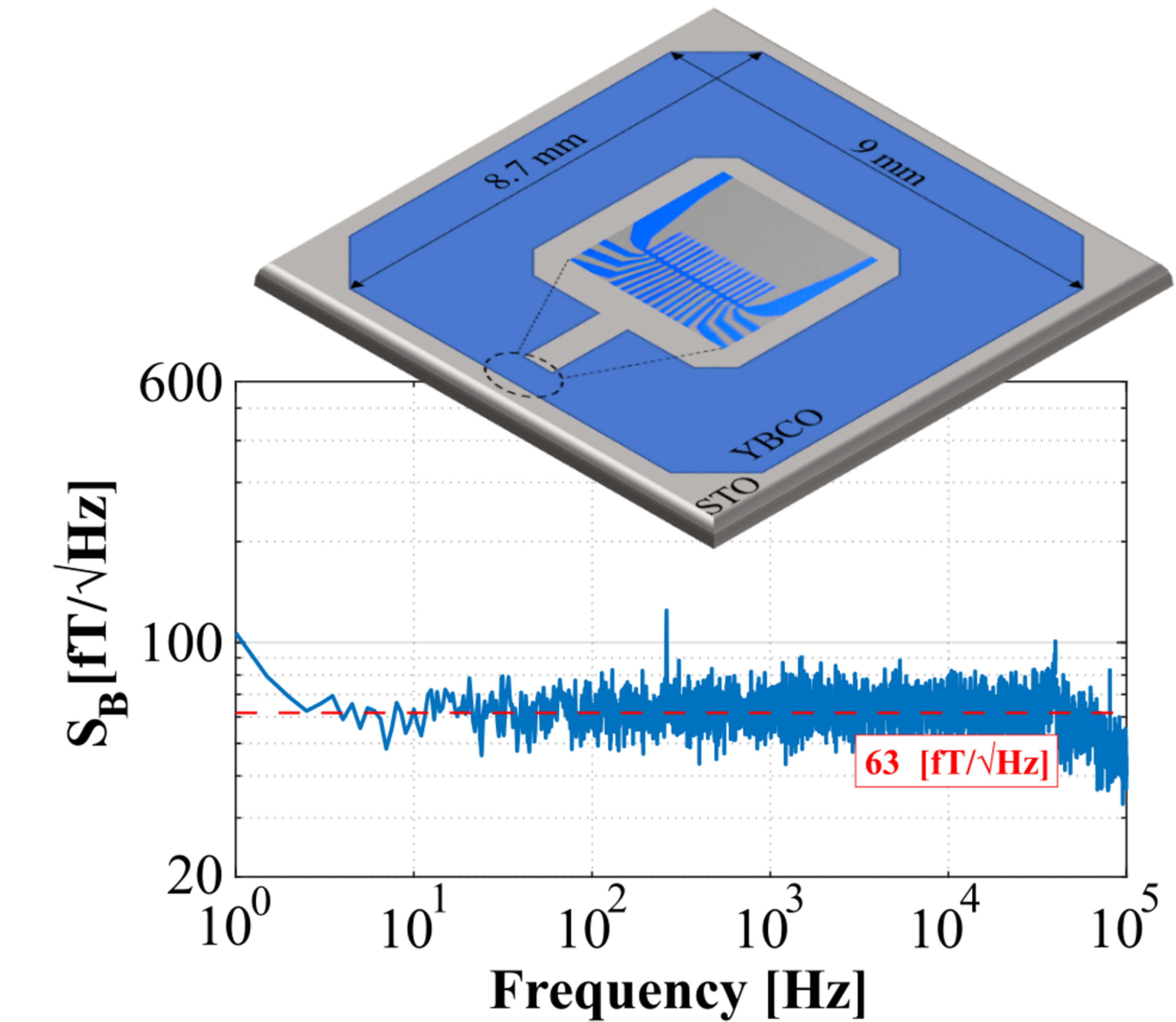}
\caption{\label{fig4} Magnetic field noise spectrum of SQ2. The inset shows the design of the pick-up loop. In the center of the loop is shown an SEM figure of a series of SQUID loops (comb-like structure) which are coupled on the sides to the pick-up loop}
\end{figure}

The magnetic field noise of SQUID-based magnetometers can be obtained from the flux noise as $S_\mathrm{B}^\mathrm{1/2}=S_\Phi^\mathrm{1/2}/A_\mathrm{eff}$, where $A_\mathrm{eff}=\Phi/B_\mathrm{a}$ (here $\Phi$ and $B_\mathrm{a}$ are the magnetic flux through the SQUID loop and the externally applied magnetic field, respectively). Due to their small size, bare SQUIDs have extremely small effective areas $A_\mathrm{eff}$, resulting in rather large values for $S_\mathrm{B}^\mathrm{1/2}$. This is a problem that is common to all SQUIDs, but particularly acute for nanoSQUIDs. To improve the magnetic field sensitivity, SQUIDs can be coupled to a much larger pick-up loop.\cite{koelle1999high} We galvanically coupled 32 SQUIDs to a single pick-up loop, which is integrated directly in the EBL design of the sample, without additional fabrication steps. A schematic of the pick-up loop is shown in the inset of Fig.~\ref{fig4}. The pick-up loop has lateral sizes of $8.7$~mm$\mathrm{x}9$~mm and line width of $2$~mm. 

An externally applied magnetic field results in a screening current circulating the loop $I_\mathrm{S}\propto B_\mathrm{a}$ which results in a phase difference between the weak links $\Delta \phi \propto I_\mathrm{S}L_\mathrm{c}2\pi/\Phi_0$. $L_\mathrm{c}$ is the SQUID hairpin loop inductance and represents the coupling inductance between SQUID and pick-up loop. This coupling determines the effective area as\cite{arzeo2016toward} $A_\mathrm{eff}=A_\mathrm{nS}+A_\mathrm{eff}^\mathrm{pl}{L_\mathrm{c}}/{L_\mathrm{loop}}$, where $A_\mathrm{nS}$ and $A_\mathrm{eff}^\mathrm{pl}$ are the effective areas of the SQUID and pick-up loop, respectively. $L_\mathrm{loop}$ is the inductance of the pick-up loop. To increase $A_\mathrm{eff}$, one needs to increase the coupling by increasing $L_\mathrm{c}$. Since a higher SQUID inductance increases the screening factor $\beta_\mathrm{L}$, in order to obtain the lowest $S_\mathrm{B}^\mathrm{1/2}$, the value of $L_\mathrm{c}$ needs to be optimized. The coupling inductance values of three SQUIDs were measured using a current injection scheme.\cite{johansson2009properties} The injection current modulates the phase difference between the two GDBs. From the modulation period of the critical current $\Delta I_\mathrm{mod}$ one can extract the inductance of the hairpin loop $L_\mathrm{c}=\Phi_0/\Delta I_\mathrm{mod}$. In Table~\ref{tab1} we summarize the values of the coupling inductance of the various SQUIDs measured at $T=77~$K. 

The magnetic field sensitivity of three different devices were measured at $T=77$~K in a magnetically shielded room. A commercial Magnicon SEL-1 dc-SQUID electronics \cite{magnicon} operated in flux-locked loop and bias reversal ($40$~kHz) mode was used to measure the magnetic flux noise. From the measured magnetic flux noise $S_\Phi^\mathrm{1/2}$, the magnetic field noise is calculated as $S_\mathrm{B}^\mathrm{1/2}=S_\Phi^\mathrm{1/2}/A_\mathrm{eff}$. Here the effective area was separately measured via responsivity measurements using a calibrated Helmholtz coil.\cite{ruffieux2020role}
\begin{table}
\caption{\label{tab1} Summary of the geometrical (gap length, $L_\mathrm{G}$, and width $W$, SQUID loop slit length $l_\mathrm{slit}$ and effective area $A_\mathrm{eff}$) and electrical properties of different SQUID-based magnetometers measured at $T=77~$K.}
\begin{ruledtabular}
\begin{tabular}{ccccccccccc}
 & $L_\mathrm{G}$ & $W$ & $l_\mathrm{slit}$ & $I_\mathrm{c}$ & $L_\mathrm{c}$ & $\beta_\mathrm{L}$ & $\Delta V_\mathrm{max}$  & $A_\mathrm{eff}$ & $S_\Phi$ & $S_\mathrm{B}$ \\ 
  & [nm] & [nm] & [$\mu$m] & [$\mu$A]& [pH] &  & [$\mu$V] & [mm$^2$] & [$\frac{\mu\Phi_0}{\sqrt{\mathrm{Hz}}}$] & [$\frac{\mathrm{fT}}{\sqrt{\mathrm{Hz}}}$]\\
\hline
SQ0 & 40 & 150 & 8 & 92 & - & - & 30 & - & -  & -\\
SQ1 & 50 & 200 & 16 & 15 & 48.5 & 0.3 & 39 & 0.15 & 6 & 85  \\
SQ2 & 50 & 200 & 30 & 30 & 103 & 1.5 & 16.5 & 0.35 & 10.6 & 63 \\
SQ3 & 50 & 200 & 30 & 16 & 103 & 0.8 & 18.7 & 0.35 & 11 & 67  \\
\end{tabular}
\end{ruledtabular}
\end{table}
In Table~\ref{tab1} we summarize the magnetic noise properties of three different SQUID magnetometers. Devices SQ2 and SQ3 were fabricated on the same chip sharing the same pick-up loop, whereas SQ1 was realized on a different chip. The best performance in terms of voltage modulations and flux noise are achieved on device SQ1. This would be expected given the shorter $l_{slit}$ and, consequently, smaller $L_\mathrm{c}$. However, this also results in a smaller $A_\mathrm{eff}$, compared to SQ2 and SQ3, due to reduced coupling. Indeed, devices SQ2 and SQ3, which have larger $l_\mathrm{slit}$, show better coupling but at the cost of a reduced $\Delta V$. Nevertheless, the $V_\Phi$ of SQ2 and SQ3 results in $S_\Phi^\mathrm{1/2}$ comparable to state-of-the-art YBCO SQUIDs ($S^\mathrm{1/2}_\Phi=2.6-10~\mu\Phi_o/\sqrt{\mathrm{Hz}}$, achieved in devices based on grain boundary junctions \cite{oisjoen2012high,faley2017high,ruffieux2020role,chukharkin2013improvement,mitchell2010ybco,faley2014graphoepitaxial}).
The lowest magnetic noise spectral density was achieved for SQ2 and is shown in Fig.~\ref{fig4}. The $1/f$ knee is below $3$~Hz and the white magnetic field noise is around $63$~fT/$\sqrt{\mathrm{Hz}}$. This result is almost a factor 2 lower than that which was previously obtained with GDB weak links\cite{trabaldo2019grooved} and represents the lowest field noise achieved in SQUID magnetometers implementing Dayem bridge based weak links. Moreover, the measured magnetic field noise is comparable with the lowest reported field noise values achieved with single-layer grain boundary-based SQUIDs galvanically coupled to a pick-up loop made on $10~$mm$\times 10$~mm substrates \cite{oisjoen2012high,ruffieux2020role,faley2014graphoepitaxial,lee1995low,beyer1998low}, $S^\mathrm{1/2}_\mathrm{B}=30-50$~fT$/\sqrt{\mathrm{Hz}}$.

In conclusion, we have studied the temperature dependence of GDB based SQUIDs. At low temperatures, the GDBs behave like short and thin Dayem bridges, whereas at temperatures above 60~K the fluctuation-driven suppression of superconductivity in the constriction of the GDB causes the weak link to behave like an SNS junction. This results in a local maximum in the temperature dependence of the transfer function around 65~K. Magnetic field noise as low as $63$~fT/$\sqrt{\mathrm{Hz}}$ at $T=77~$K has been obtained for  GDB based magnetometers galvanically coupled to an in-plane pick-up loop. 
Such low values makes these devices an attractive candidate for magnetoencephalography application, possibly outperforming their LTS counterpart.\cite{riaz2017evaluation}

\begin{acknowledgments}
This work  was  been  supported  in  part  by  the  Knut  and  Alice Wallenberg  Foundation  (KAW)  and  in part  by  the  Swedish Research Council (VR). This project has received funding from the  ATTRACT  project  funded  by  the  EC  under  Grant Agreement 777222. R.A. is  supported  by  the  Swedish Research  Council  (VR)  under  the  project  2017-00382. This work was performed in part at Myfab Chalmers.
\end{acknowledgments}

\bibliography{bibliography}

\end{document}